\documentstyle[12pt]{article}
\newcommand{\bec}{\begin{center}}
\newcommand{\ec}{\end{center}}
\newcommand{\gw}{Gromov-Witten}
\newcommand{\cy}{Calabi-Yau}
\newcommand{\bee}{\begin{equation}}
\newcommand{\ee}{\end{equation}}
\begin{document}
\large
\begin{titlepage}
\bec
{\Large\bf  Matter in F-Theory\\}
\vspace*{15mm}
{\bf Yu. Malyuta and T. Obikhod\\}
\vspace*{10mm}
{\it Institute for Nuclear Research\\
National Academy of
Sciences of Ukraine\\
252022 Kiev, Ukraine\\}
{\bf e-mail: interdep@kinr.kiev.ua\\}
\vspace*{35mm}
{\bf Abstract\\}
\ec
\vspace*{1mm}
It is shown that the matter content of F-theory
compactifications on elliptic \cy\ threefolds
is encoded in the \gw\ invariants.
\end{titlepage}
\section {\bf Introduction}
The matter content of six-dimensional theories
resulting from compactification of F-theory
on elliptic \cy\ threefolds has been
obtained in \cite{1.} by using Tate's algorithm.\\
\hspace*{6mm}Another algorithm for obtaining the 
matter content of six-dimensional theories has
been proposed in \cite{2.}. This algorithm
allows to obtain the matter content from the
generalized Green-Schwarz anomaly cancellation
condition. \\
\hspace*{6mm}The purpose of the present paper
is to show that the matter content of F-theory
compactifications on elliptic \cy\ threefolds
is encoded in the \gw\ invariants.
\section {\bf \gw\ invariants}
Let us consider the chains of elliptic fibred
\cy\ threefolds recorded in Table 1, where 
$ F_{n} $ is the Hirzebruch surface and 
$ (h^{21}, h^{11}) $ are the Hodge numbers.
\bec
\scriptsize
\begin{tabular}{|c|llllllll|} \hline
 & & & & & & & & \\
{\footnotesize $ F_{n} $\/} & \hspace*{0.5mm} 
{\footnotesize $ SU(1) $\/}  & 
\hspace*{0.5mm} {\footnotesize $ SU(2) $\/}  & 
\hspace*{0.5mm} {\footnotesize $ SU(3) $\/}  & 
\hspace*{0.5mm} {\footnotesize $ SU(4) $\/}  & 
\hspace*{0.5mm} {\footnotesize $ SU(5) $\/}  & 
{\footnotesize $ SO(10) $\/}  & 
\hspace*{3mm}   {\footnotesize $ E_{6} $\/}  & 
\hspace*{3mm}   {\footnotesize $ E_{7} $\/}   \\
 & & & & & & & & \\ \hline \hline
 & & & & & & & & \\
{\footnotesize $ F_{2} $\/} &
(243, 3)&(190, 4)&(161, 5) & 
(138, 6)  & (117, 7) & 
(112, 8) & (105, 9) & (96, 10) \\
 & & & & & & & & \\
{\footnotesize $ F_{4} $\/}  & 
(271, 7) & (194, 8) & 
(153, 9) & (122, 10) & (95, 11) & 
(88, 12) & (79, 13) & (68, 14)  \\
 & & & & & & & & \\
{\footnotesize $ F_{6} $\/}  & 
(321, 9) & (220, 10) & 
(167, 11) & (128, 12) & (95, 13) & 
(86, 14) & (75, 15) & (62, 16) \\
 & & & & & & & & \\
{\footnotesize $ F_{8} $\/}  & 
(376, 10) & (251, 11) & 
(186, 12) & (139, 13) & (100, 14) & 
(89, 15) & (76, 16) & (61, 17) \\
 & & & & & & & & \\
{\footnotesize $ F_{10} $\/}  & 
(433, 13) & (284, 14) & 
(207, 15) & (152, 16) & (107, 17) & 
(94, 18) & (79, 19) & (62, 20) \\
 & & & & & & & & \\
{\footnotesize $ F_{12} $\/}  & 
(491, 11) & (318, 12) & 
(229, 13) & (166, 14) & (115, 15) & 
(100, 16) & (83, 17) & (64, 18) \\ 
 & & & & & & & & \\ \hline
\end{tabular}\\
\large
\vspace*{5mm}
{\bf Table 1 :} \ The chains of elliptic 
fibred \cy\ threefolds.
\ec
\newpage
\large
\hspace*{6mm} The chains for each $ n $ have 
similar properties, therefore we will treat 
in some detail only the first three models 
of the chain $ n=2 $. Application of the 
program INSTANTON \cite{3.}
to these models gives the \gw\ invariants 
recorded in Tables 2, 3 and 4. Note that our 
normalization is given by 
\[ \frac{1}{2}J^{3} , \]
where $ J^{3} $ is the topological coupling 
\cite{4.}.
\small
\bec
\vspace*{10mm}
\begin{tabular}{|lrlrlrlr|}      \hline
(0,0,1) &   -1 & (0,1,1) &   -1  & 
(0,1,2) &   -2 & (0,1,3) &  -3  \\
(0,1,4) &   -4 & (0,1,5) &   -5  & 
(0,2,3) &   -3 & (0,2,4) & -16  \\
(1,0,0) &  240 & (1,0,1) &  240  & 
(1,1,1) &  240 & (1,1,2) & 720  \\
(1,1,3) & 1200 & (1,1,4) & 1680  & 
(1,2,3) & 1200 & (2,0,0) & 240  \\
(2,0,2) &  240 & (2,2,2) &  240  & 
(3,0,0) &  240 & (3,0,3) & 240  \\
(4,0,0) &  240 & (5,0,0) &  240  & 
(6,0,0) &  240 & (0,1,0) &    0  \\ \hline
\end{tabular}\\
\vspace*{10mm}
\large
{\bf Table 2 :} The \gw\ invariants for 
{$ X_{24}(1,1,2,8,12)^{3,243}_{-480} $} \ . \\
\vspace*{10mm}
\small
\begin{tabular}{|lrlrlrlr|}      \hline
(0,0,0,1) &   28 & (0,0,0,2)&   -1  & 
(0,0,0,3) &  0 & (0,0,1,0) &  -1  \\
(0,1,0,0) &   0 & (0,1,1,0) &   -1  
& (0,1,2,0) & -2 & (0,1,3,0) &  -3  \\
(0,1,4,0) &  -4 & (0,1,5,0) &   -5  
& (0,2,3,0) & -3 & (0,2,4,0) & -16  \\
(0,2,5,0) & -55 & (0,2,6,0) & -144  
& (0,3,4,0) & -4 & (0,3,4,0) &  -4  \\
(0,3,5,0) & -55 & (1,0,0,0) &   -1  
& (1,0,0,1) & 28 & (1,0,0,2) & 186  \\
(1,0,0,3) &  28 & (1,0,0,4) &   -1  
&           &     &          &       \\ \hline
\end{tabular}\\
\vspace*{10mm}
\large
{\bf Table 3 :} The \gw\ invariants for 
{$  X_{20}(1,1,2,6,10)^{4,190}_{-372} $} \ . \\
\vspace*{10mm}
\small
\begin{tabular}{|lrlrlrlr|}      \hline
(0,0,0,0,1) &   15 & (0,0,0,0,2) & 0 
& (0,0,0,1,0) & -1 & (0,0,0,1,1) & 15  \\
(0,0,0,1,2) &   15 & (0,0,0,1,3) & -1 
& (0,0,1,0,0) & -1 & (0,1,0,0,0) & 0  \\
(0,1,1,0,0) &   -1 & (0,1,2,0,0) & -2 
& (0,1,3,0,0) & -3 & (0,1,4,0,0) & -4  \\
(0,2,3,0,0) &   -3 & (0,2,4,0,0) & -16 
& (0,2,5,0,0) & -55 & (1,0,0,0,0) & -1  \\
(1,0,0,1,0) &   -1 & (1,0,0,1,1) & 15 
& (1,0,0,1,2) & 15 & (1,0,0,1,3) & -1  \\
(1,0,0,2,2) &   15 & (1,0,0,2,3) & 156 
& (1,0,0,2,4) & 15 & (1,0,1,0,0) & -1  \\
(1,0,0,3,0) &   -1 & (1,0,0,3,1) & 15 
& (1,0,0,3,2) & 15 & (1,0,0,3,3) & -1  \\ \hline
\end{tabular}\\
\vspace*{10mm}
\large
{\bf Table 4 :} \ The \gw\ invariants for 
{$  X_{18}(1,1,2,6,8)^{5,161}_{-312} $} \ . \\
\vspace*{8mm}
\ec
\large
\section{\bf Matter content}
From Tables 2 and 3 we infer the relation for the 
\gw\ invariants 
\bee
n_{a,b,c}=\sum_{k}n_{a,b,c,k} \ .
\ee
This relation describes the phase transition 
between the models  
$ X_{24}(1,1,2,8,12)^{3,243}_{-480} $
and $  X_{20}(1,1,2,6,10)^{4,190}_{-372} $ \ .
Let us consider the following realization of (1)
\bee
240 = -1 + 28 + 186 +28 - 1\ .
\ee
The matter content of the model 
$  X_{20}(1,1,2,6,10)^{4,190}_{-372} $
can be read off from (2). The result is
\[ 28 \ \ {\bf 2}\ . \]
\hspace*{6mm}From Tables 3 and 4 we infer 
the relation for the \gw\ invariants
\bee
n_{a,b,c,d}=\sum_{k}n_{a,b,c,d,k}\ .
\ee
This relation describes the phase transition
between the models 
$  X_{20}(1,1,2,6,10)^{4,190}_{-372} $
and $  X_{18}(1,1,2,6,8)^{5,161}_{-312} $\ .
Let us consider the following realizations of (3)
\bee
\begin{tabular}{rrrrrrrr}
 28 & = & $-1 $ & + & 15 & + & 15 & $ -1 $\ , \\
   186 & = & 15 & + & 156 & + & 15 & \hspace*{4mm} , \\
   28 & = & $ -1 $ & + & 15 & + & 15 & $ -1 $\ . \\
\end{tabular}
\ee
The matter content of the model 
$  X_{18}(1,1,2,6,8)^{5,161}_{-312} $
can be read off from (4). The result is
\[30 \  \ {\bf 3}\ . \]
\hspace*{6mm}Applying the analogous 
consideration to the remaining models
listed in Table 1 we obtain the matter 
content recorded in Table 5.
\bec
\begin{tabular}{|c|c|}  \hline
                  &              \\
Group             &  Matter content  \\
                  &         \\ \hline  \hline
                  &            \\
$ SU(2) $         & $   (6n+16){\bf 2}  $\\
                  &              \\
$ SU(3) $         & $   (6n+18){\bf 3}  $ \\
                  &               \\
$ SU(4) $         
& $   (n+2){\bf 6}+(4n+16){\bf 4}  $ \\
                  &              \\
$ SU(5) $         
& $   (3n+16){\bf 5}+(2+n){\bf 10} $ \\
                  &             \\
$ SO(10)$         
& $   (n+4){\bf 16}+(n+6){\bf 10} $ \\
                  &             \\
$ E_{6} $         & $ (n+6){\bf 27}$  \\
                  &              \\
$ E_{7} $         
& $ (\frac{n}{2}+4){\bf 56} $  \\
                  &           \\    \hline
\end{tabular}\\
\vspace*{10mm}
{\bf Table 5 :} Matter content of models with 
enhanced gauge groups  (for $ n=2, 4, 6, 8, 10, 12 $) .
\ec
\vspace*{15mm}
{\bf Acknowledgement}
\vspace*{5mm}\\
We wish to thank Professor O.S. Parasyuk
for useful discussions.

\end{document}